\documentstyle[psfig]{mn}
\newif\ifAMStwofonts
\newcommand{\gapp}{\mbox{\raisebox{-0.3em}{$\stackrel{\textstyle >}{\sim}$}}}

\title[A multifrequency study of possible relic lobes]
{A multifrequency study of possible relic lobes in giant radio sources} 
\author[Godambe et al.]
       {Sagar Godambe$^{1,2}$$\thanks{E-mail: sagar.godambe@gmail.com (SG); chiranjib@iucaa.ernet.in (CK);
                         djs@ncra.tifr.res.in (DJS); wiita@chara.gsu.edu (PJW)}$, C. Konar$^{3}$, 
        D.J. Saikia$^{1}$ and Paul J. Wiita$^{4,5}$ \\
$^{1}$ National Centre for Radio Astrophysics, Tata Institute of Fundamental Research, Post Bag 3,
       Ganeshkhind, Pune 411007, India  \\
$^{2}$ Department of Physics, The University of Utah, Salt Lake City, UT 84112-0830, USA \\
$^{3}$ Inter-University Centre for Astronomy and Astrophysics, Post Bag 4,
       Ganeshkhind, Pune 411007, India  \\
$^{4}$ Department of Physics and Astronomy, Georgia State University, PO Box 4106, Atlanta, Georgia 30302-4106, USA \\ 
$^{5}$ School of Natural Sciences, Institute for Advanced Study, Princeton, New Jersey, 
       08540, USA \\   
       }
\date{Accepted.    Received }

\pagerange{\pageref{firstpage}--\pageref{lastpage}}
\pubyear{  }

\begin{document}

\maketitle

\label{firstpage}

\begin{abstract}
We present low-frequency observations with the Giant Metrewave Radio Telescope (GMRT) of 
three giant radio sources (GRSs, J0139+3957, J0200+4049 and J0807+7400) with relaxed diffuse 
lobes which show no hotspots and no evidence of jets. The largest of these three, 
J0200+4049, exhibits a depression in the centre of the western lobe, while 
J0139+3957 and  J0807+7400 have been suggested earlier by Klein et al. and Lara et al. 
respectively to be relic radio sources.  We estimate the ages
of the lobes. We also present Very Large Array (VLA) observations of 
the core of J0807+7400, and determine the core radio spectra for all three sources.  
Although the radio cores suggest that the sources are currently active, we explore the 
possibility that the lobes in these sources are due to an earlier cycle of activity. 
\end{abstract}

\begin{keywords}
galaxies: active -- galaxies: individual (J0139+3957, J0200+4049 and J0807+7400) -- 
galaxies: nuclei -- radio continuum: galaxies
\end{keywords}

\section{Introduction}
Giant radio sources (GRSs) are defined to be those which have a projected linear size
$\gapp$1 Mpc (H$_o$=71 km s$^{-1}$ Mpc$^{-1}$, $\Omega_m$=0.27,
$\Omega_{vac}$=0.73, Spergel et al. 2003), and provide useful insights towards
understanding the late stages of evolution of radio sources and probing 
the external environment  on Mpc scales at different redshifts
(e.g. Gopal-Krishna, Wiita \& Saripalli 1989; 
Subrahmanyan \& Saripalli 1993; Subrahmanyan, Saripalli \& Hunstead 1996;
Mack et al. 1998; Ishwara-Chandra \& Saikia 1999;
Kaiser \& Alexander 1999; Blundell, Rawlings \& Willott 1999 and references therein; 
Schoenmakers et al. 2000, 2001; Konar et al. 2004, 2008; Machalski et al. 2007, 2008;
Jamrozy et al. 2008).

There have been many attempts to determine the ages of double-lobed radio sources
including the GRSs.  Most lobes of GRSs exhibit a steepening in spectral index from the 
outer peaks towards the nuclear region, allowing an estimate of the spectral age, similar to
what has been done for double-lobed radio sources of smaller dimensions. For a 
sample of GRSs observed with the Giant Metrewave Radio Telescope (GMRT) and the
Very Large Array (VLA), Jamrozy et al. (2008) have estimated their median age to be 
$\sim$20 Myr while their median size is $\sim$1300 kpc. The spectral age is likely 
to be a lower limit since the radio emission is often not seen all the way to the 
nuclear region, and there could also be re-acceleration of particles in the lobes.  
For the smaller sources a large number of studies of spectral ages have been made. 
For example, for a sample of 
3CR sources with a median size of 342 kpc studied by Leahy, Muxlow \& Stephens (1989) using 
the Multi-Element Radio Linked Interferometer Network (MERLIN) at 151 MHz and
the VLA at 1500 MHz, the spectral age was found to have a median value of $\sim$8 Myr.
Similar gradients in spectral index across the lobes have also been seen in
smaller sources such as those studied by Liu, Pooley \& Riley (1992) which have a median
linear size of 103 kpc and a median age of $\sim$1.7 Myr. There is a clear trend   
for the spectral age to increase with size, which is broadly consistent with
the expectations of dynamical models of the propagation of jets in an external
medium (e.g. Falle 1991; Kaiser \& Alexander 1997; Jeyakumar et al. 2005 and
references therein). 

In the course of our study of GRSs, a number of these objects appeared to have
diffuse lobes of emission without any prominent peaks of emission towards the
outer edges of the lobes. Unlike Fanaroff-Riley Class I sources, these do not
have any jets and the radio emission could be from relic lobes. 
We have chosen three of these sources, namely 
J0139+3957, J0200+4049 and J0807+7400, for a detailed
investigation using the GMRT.  Two of these, J0139+3957 and J0807+7400, have
been suggested earlier to be
relic radio sources by Klein et al. (1995) and Lara et al. (2001) respectively. 
In this paper we present the 
results of multifrequency observations of these three sources at low 
frequencies with the GMRT, attempt to 
estimate their spectral ages and explore the possibility that these lobes 
might be relics from an earlier cycle of activity. We also present the core
flux densities of J0807+7400 from archival VLA data, and the core spectra
of the three sources.
 
%%%%%%%%%%%%%%%%%%%%%%%%%%%%%%%%%%%%%%%%%%%%%%
\begin{table}
\caption{ Observing log }
\begin{tabular}{l c c c c }

\hline
Teles-    & Array  & Obs.   &  Sources               & Obs.       \\
cope      & Conf.  & Freq.  &                        & Date       \\
          &        & MHz    &                        &            \\
  (1)     &  (2)   & (3)    &   (4)                  & (5)        \\
\hline
GMRT      &        & 239    & J0139+3957             & 2007 Jun 02 \\
GMRT      &        & 334    & J0139+3957             & 2005 Jan 28 \\
GMRT      &        & 605    & J0139+3957             & 2007 Jun 02 \\
GMRT      &        & 1287   & J0139+3957             & 2003 Aug 19 \\
VLA$^{a}$ &   D    & 4841   & J0139+3957             & 2000 Jul 24 \\
%-----------------------------------------------------------------------
GMRT      &        & 239    & J0200+4049             & 2007 Dec 29 \\
GMRT      &        & 333    & J0200+4049             & 2004 Dec 25 \\
GMRT      &        & 605    & J0200+4049             & 2007 Dec 29 \\
GMRT      &        & 1289   & J0200+4049             & 2004 Nov 25 \\
%---------------------------------------------------------------------
GMRT      &        & 239    & J0807+7400             & 2005 Jan 07 \\     
GMRT      &        & 334    & J0807+7400             & 2004 Dec 07 \\     
GMRT      &        & 605    & J0807+7400             & 2005 Jan 07 \\     
GMRT      &        & 1289   & J0807+7400             & 2005 Jan 17 \\     
%%%%%%%%%%
VLA$^{b}$ & B      & 1385   & J0807+7400             & 1995 Nov 19 \\
VLA$^{b}$ & D      & 1435   & J0807+7400             & 1993 Nov 01 \\
VLA$^{b}$ & C      & 1685   & J0807+7400             & 1996 Feb 19 \\
VLA$^{b}$ & C      & 4860   & J0807+7400             & 1996 Feb 19 \\
%---------------------------------------------------------------------
\hline
\end{tabular}

$^a$ Image published by Konar et al. (2004). \\
$^b$ Archival data from the VLA. \\
\end{table}
%%%%%%%%%%%%%%%%%%%%%%%%%%%%%%%%%%%%%%%%%%%%%%

\section{Observations and analyses} 
Both the GMRT and the VLA observations were made 
in the standard fashion, with each target source 
observation interspersed with observations of the phase calibrator. The primary 
flux density calibrator was any one of 3C48, 3C147 and 3C286 with all flux 
densities being on the scale of Baars et al. (1977). 
The total observing time on each source is a few hours for the 
GMRT observations while for the VLA observations the time on source
ranges up to 10 minutes.  All the data were analysed in the standard 
fashion using the NRAO {\tt AIPS} package. All the data were self calibrated 
to produce the best possible images.

The observing log for both the GMRT and the VLA observations is given in 
Table 1 which is arranged as follows. Columns 1 and 2 show the name of the 
telescope and the array configuration for the VLA observations;
column 3 shows the frequency of the observations in MHz, while 
columns 4 and 5 list the sources observed and the dates of the observations, 
respectively. 

\section{Observational results}
The GMRT images of the three sources, J0139+3957, J0200+4049 and J0807+7400,
 at the different frequencies are presented in
Figs. 1, 2 and 3 respectively,  while the observational parameters and some of the observed
properties are presented in Table 2, which is arranged as follows.
Column 1: Name of the source; column 2: frequency  of observations in units of MHz; 
columns 3-5: the major and minor axes of the restoring beam in arcsec and its position angle 
(PA) in degrees;
column 6: the rms noise in units of mJy beam$^{-1}$; column 7: the integrated flux density of the
source in mJy estimated by specifying an area enclosing the entire source. We examined the
change in flux density by specifying different areas and found the difference to be within
a few per cent. The flux densities at different frequencies have been estimated over 
similar areas.  Columns 8, 11 and 14: component designation, where W, C and E denote 
the western, core and eastern components respectively;
columns 9 and 10, 12 and 13, and 15 and 16: the peak and total flux densities of each of the
components in units of mJy beam$^{-1}$ and mJy respectively.  The superscript $g$ 
indicates that the flux densities have been estimated from a two-dimensional Gaussian fit to the 
core component. 

%%%%%%%%%%%%%%%%%%%%%%%%%%%%%%%%%%%%%%%%%%%%%%%%%%%%%%%%%%%%%%%%%%%%%%%%%
\begin{figure*}
\vbox{
  \hbox{
  \psfig{file=J0139_T.CONT.PS,width=3.45in,angle=-90}
  \psfig{file=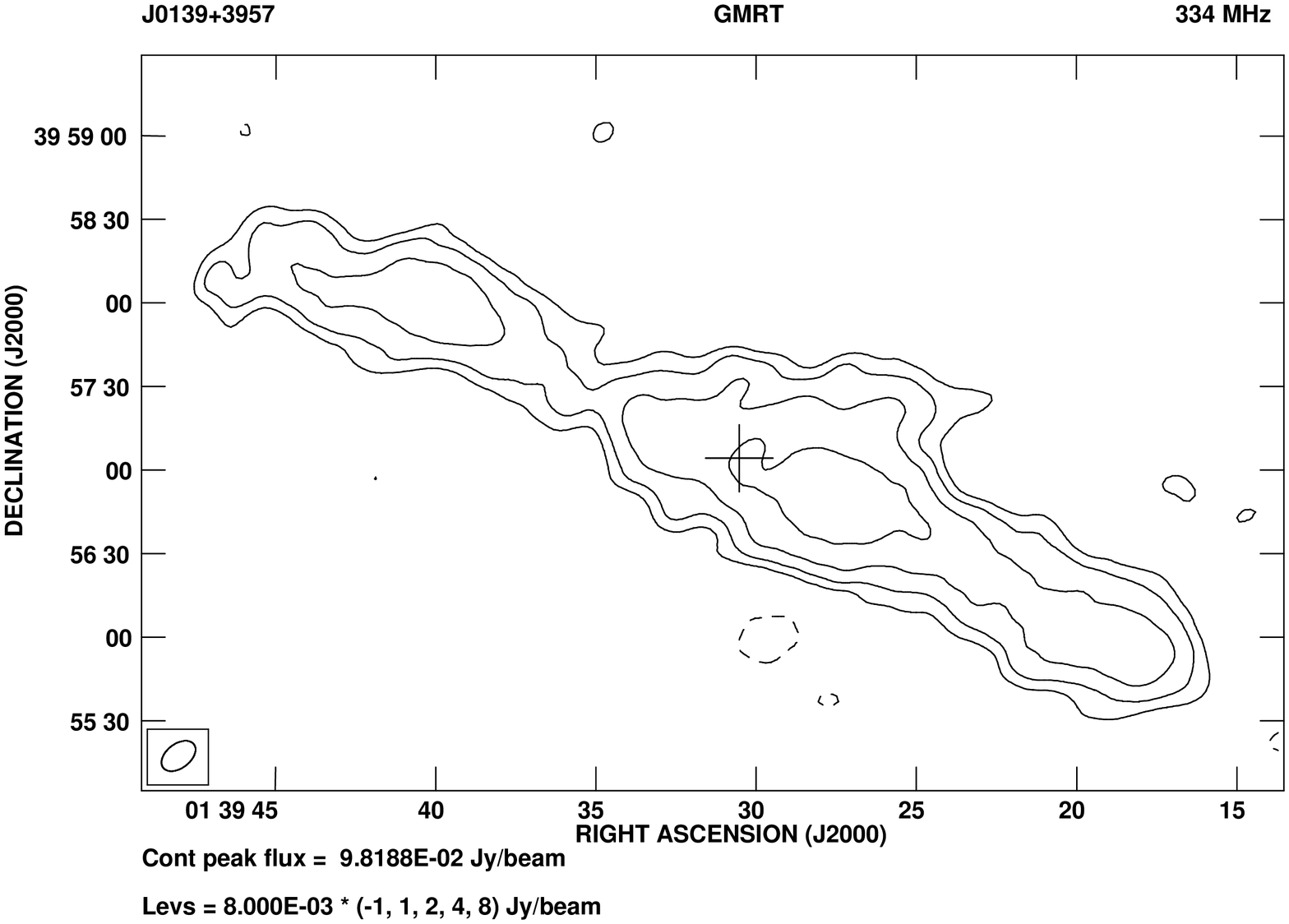,width=3.45in,angle=-90}
       }
  \hbox{
  \psfig{file=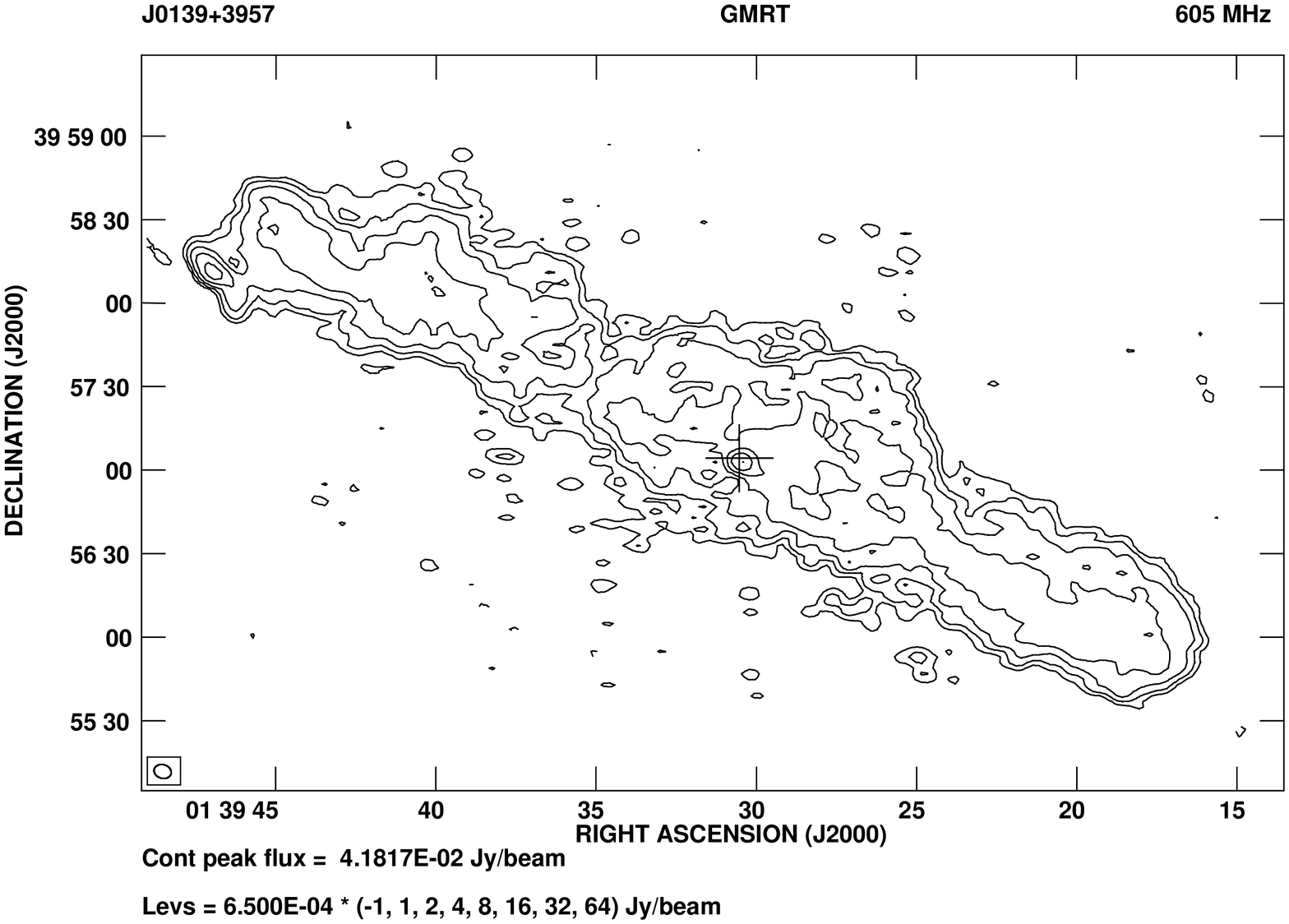,width=3.45in,angle=-90}
  \psfig{file=J0139_L.CONT.PS,width=3.45in,angle=-90}
       }
}
\caption[]{GMRT images of J0139+3957 at 239, 334, 605 and 1287 MHz. The image
at 1287 MHz has been reproduced from Konar et al. (2004). In this
figure as well as in all the other images of the sources, the peak brightness and
the contour levels in units of Jy beam$^{-1}$ are given below each image. 
In all the images the restoring beams are indicated by ellipses; the values are
those listed in Table 2 unless mentioned otherwise in the caption.
The + sign indicates the position of the optical host galaxy.
           }
\end{figure*}
%%%%%%%%%%%%%%%%%%%%%%%%%%%%%%%%%%%%%%%%%%%%%%%%%%%%%%%%%%%%%%%%%%%%%%%%%

%%%%%%%%%%%%%%%%%%%%%%%%%%%%%%%%%%%%%%%%%%%%%%%%%%%%%%%%%%%%%%%%%%%%%%%%%
\begin{figure*}
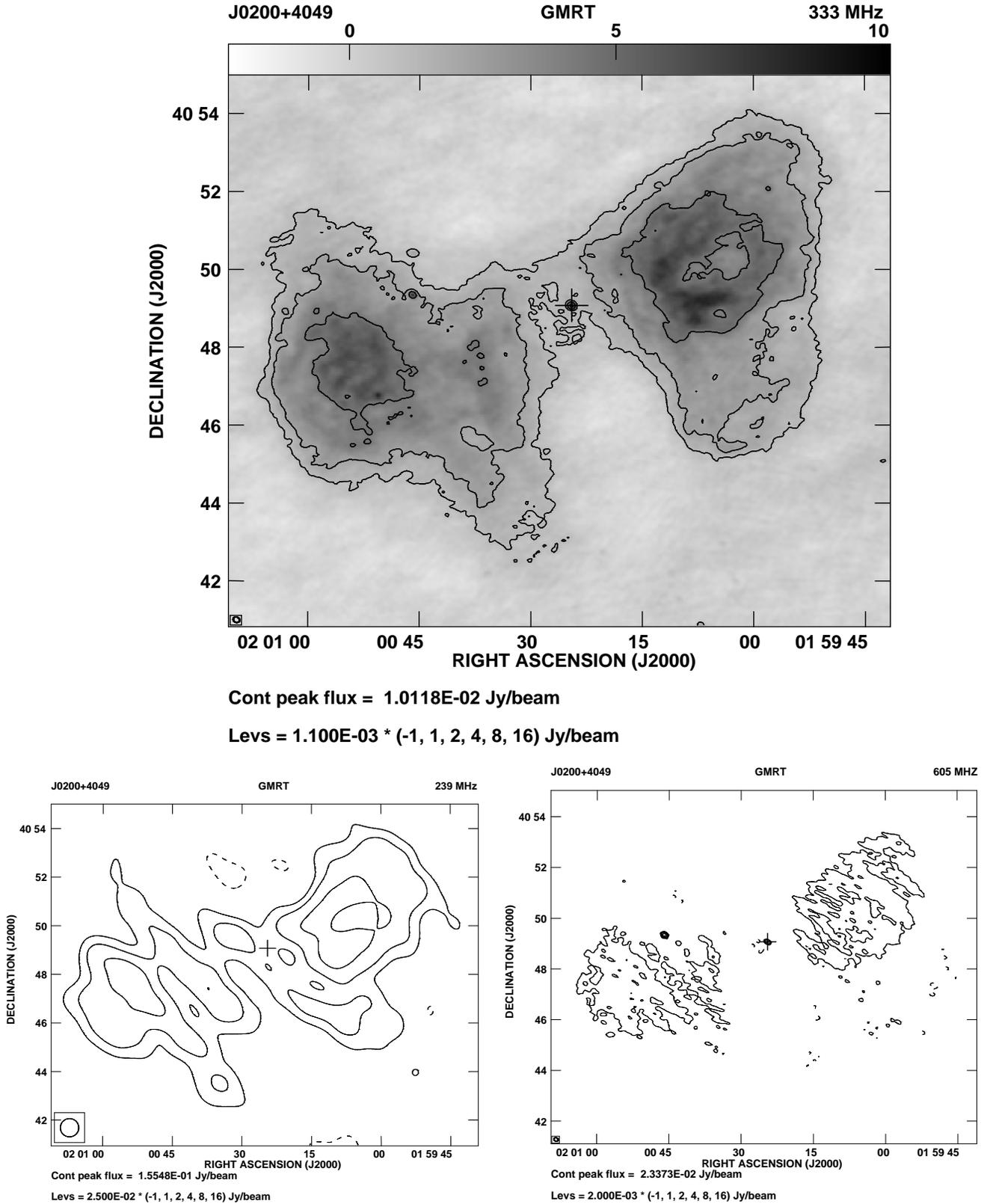

\vbox{
  \hspace{1.0cm}
  \hbox{
  \psfig{file=J0200P_GREY.PS,width=5.45in,angle=0}
       }
  \hbox{
  \psfig{file=J0200_T.CONV45.PS,width=3.45in,angle=-90}
  \psfig{file=J0200_G.CONVL.PS,width=3.45in,angle=-90}
       }
}
\caption[]{GMRT image of J0200+4049 at 333 MHz is shown in the upper panel.
The 239-MHz image convolved to a resolution of 45 arcsec, and the 605-MHz 
image convolved to the resolution of the 333-MHz image are shown in the
lower panels.
           }
\end{figure*}
%%%%%%%%%%%%%%%%%%%%%%%%%%%%%%%%%%%%%%%%%%%%%%%%%%%%%%%%%%%%%%%%%%%%%%%%%

%%%%%%%%%%%%%%%%%%%%%%%%%%%%%%%%%%%%%%%%%%%%%%%%%%%%%%%%%%%%%%%%%%%%%%%%%
\begin{center}
\begin{figure*}
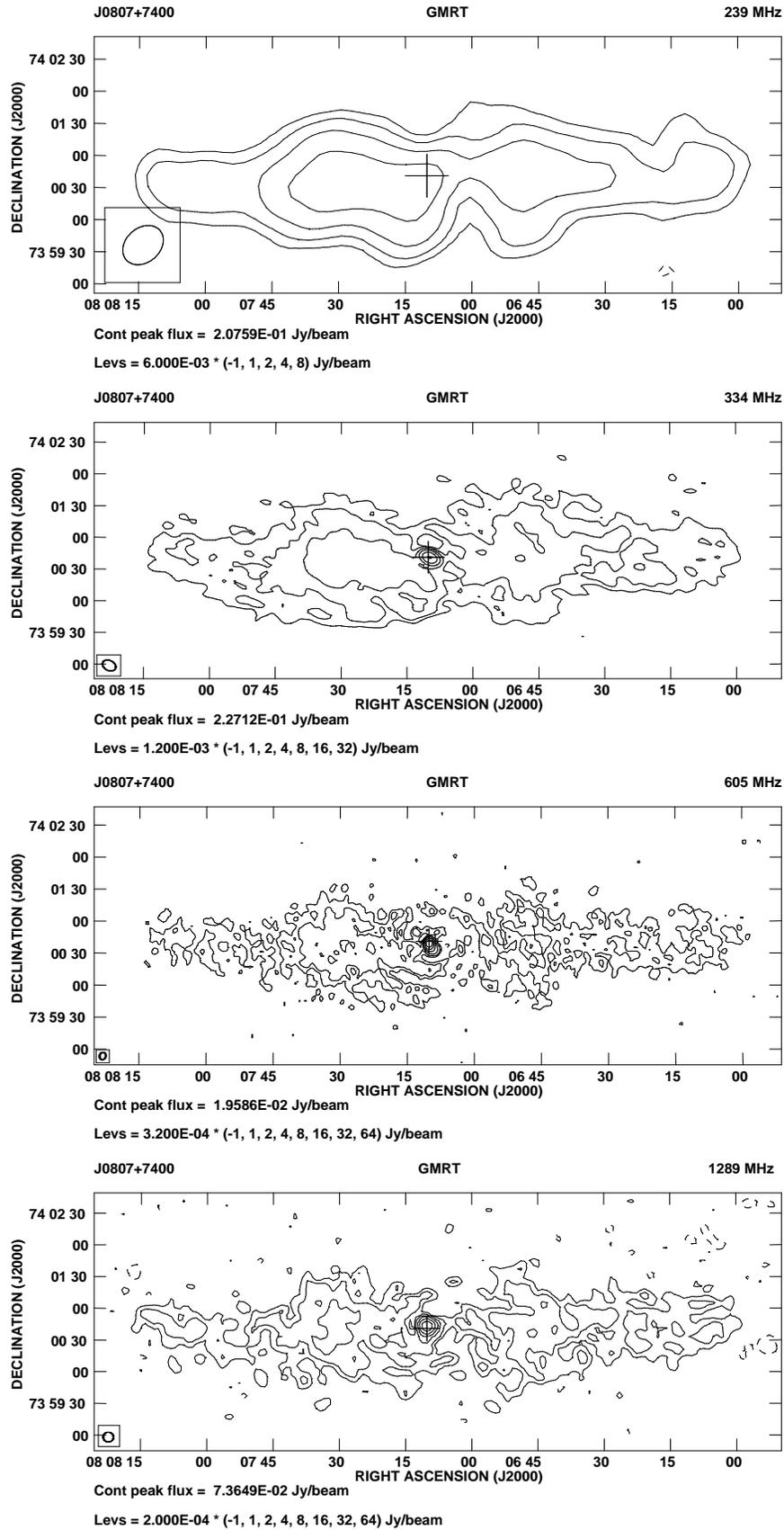

\hspace{2.0cm}
\vbox{
  \psfig{file=J0807_T.CONT.PS,width=4.65in,angle=-90}
  \psfig{file=J0807_P.CONT.PS,width=4.65in,angle=-90}
  \psfig{file=J0807_G.CONT.PS,width=4.65in,angle=-90}
  \psfig{file=J0807_L.CONT.PS,width=4.65in,angle=-90}
}
\caption[]{GMRT images of J0807+7400 at 239, 334, 605 and 1289 MHz.
           }
\end{figure*}
\end{center}
%%%%%%%%%%%%%%%%%%%%%%%%%%%%%%%%%%%%%%%%%%%%%%%%%%%%%%%%%%%%%%%%%%%%%%%%%

%%%%%%%%%%%%%%%%%%%%%%%%%%%%%%%%%%%%%%%%%%%%%%%%%%%%%%%%%%%%%%%%%%%%%%%%%
\begin{table*}
\caption{ The observational parameters and observed properties of the sources from the GMRT images.}

\begin{tabular}{l l rrr r r r rr l rr r rr}
\hline
Source   & Freq.       & \multicolumn{3}{c}{Beam size}                    & rms      & S$_I$   & Cp  & S$_p$  & S$_t$  & Cp   & S$_p$ & S$_t$ & Cp  & S$_p$   & S$_t$     \\

           & MHz         & $^{\prime\prime}$ & $^{\prime\prime}$ & $^\circ$ &    mJy   & mJy     &     & mJy    & mJy    &      & mJy   & mJy   &     & mJy     & mJy       \\
%           &             &                   &                   &          & beam$^{-1}$&       &     &beam$^{-1}$   &        &      & beam$^{-1}$    &       &     & beam$^{-1}$      &           \\ 
            &             &                   &                   &          &      /b    &       &     &     /b      &        &      &    /b         &       &     &    /b           &           \\ 
   (1)     & (2)   & (3)  & (4)  & (5)  & (6)  & (7)  &(8)& (9)  & (10) & (11)  &   (12)  &(13)  &(14)& (15) & (16)  \\
\hline
J0139+3957 &  239  & 14.3 & 12.9 &  278 & 2.85 & 6415 & W & 211  & 3609 &      &           &     & E & 143 & 2728   \\
           &  334  & 13.9 & 8.6  &  308 & 2.58 & 5434 & W & 103  & 3403 &      &           &     & E & 75  & 2044   \\
           &  605  & 6.3  & 4.9  &  74  & 0.16 & 2643 & W & 42   & 1445 &C$^g$ & 38        & 74  & E & 42  & 1249   \\ 
           &  1287 & 6.2  & 3.8  &  37  & 0.05 & 1201 & W & 4.5  & 600  &C$^g$ & 34        & 35  & E & 10  & 579    \\
%---------------------------------------------------------------------------------------------------------------------
J0200+4049 & 239   & 20.5 & 12.6 & 58   & 0.83 & 9129 & W & 25  & 4788  &      &           &     & E & 25  & 4543   \\
           & 333   & 10.2 & 8.2  & 68   & 0.28 & 10000& W & 8.4 & 5028  &C$^g$ & 9.5       & 14.8& E & 8.1 & 5183   \\
           & 605   & 5.4  & 5.3  & 88   & 0.13 & 5014 & W & 2.3 & 2733  &C$^g$ & 8.4       & 9.6 & E & 2.5 & 2384   \\
           & 1289  & 3.7  & 2.5  & 316  & 0.12 &      &   &     &       &C     & 6.4       & 8.9 &   &     &        \\  
%---------------------------------------------------------------------------------------------------------------------
J0807+7400 & 239   & 41.8 & 32.0 & 312  & 1.5  & 929  & W & 74  & 404   &      &           &     & E & 80  & 557    \\
           & 334   & 13.2 & 9.5  & 63   & 0.32 & 937  & W & 6.2 & 396   &C$^g$ & 40        & 45  & E & 8.8 & 493    \\
           & 605   & 7.0  & 5.7  & 346  & 0.08 & 433  & W & 14  & 192   &C$^g$ & 19        & 22  & E & 20  & 238    \\ 
           & 1289  & 9.6  & 9.3  & 83   & 0.05 & 169  & W & 11  & 81    &C$^g$ & 8.5       & 11.6& E & 7.4 & 87     \\
%---------------------------------------------------------------------------------------------------------------------
\hline
\end{tabular}

C: Values of core flux densities listed here have been estimated from images with the resolutions given here. Estimates 
   by mapping with a lower uv-cutoff are listed later, along with values from the literature. 
\end{table*}
%%%%%%%%%%%%%%%%%%%%%%%%%%%%%%%%%%%%%%%%%%%%%%%%%%%%%%%%%%%%%%%%%%%%%%%%%%%%%%%%%%%%%%%%%%%%%%%%%%%%%%%%%%%%%%%%%%%%%%%%%%%

%%%%%%%%%%%%%%%%%%%%%%%%%%%%%%%%%%%%%%%%%%%%%%%%%%%%%%%%%%%%%%%%%%%%%%%%%
\begin{table}
\caption{The total flux densities}

\begin{tabular}{l l r r r} 
\hline
Source   & Freq.      & Flux & Error & References  \\
         &            & density &    &             \\
         & MHz        &  mJy  & mJy  &             \\
   (1)   & (2)        & (3)   & (4)  & (5)         \\  
\hline
J0139+3957 & 26.3      & 64960 &6086  &   9  \\ 
           & 74        & 17821 &2673  &   1   \\ 
           & 151       & 10800 & 756  &   10  \\ 
           & 178       & 4773  & 596  &   11  \\ 
           & 239       & 6415  & 962  &   P   \\
           & 325       & 4750  & 100  &   15  \\  
           & 334       & 5434  & 815  &   P   \\
           & 408       & 3890  & 195  &   12  \\ 
           & 605       & 2643  & 264  &   P   \\       
           & 1287      & 1201  & 120  &   P   \\
           & 1400      & 798   &  16  &   15  \\  
           & 1400      & 802   &  80  &   2   \\       
           & 1460      & 1037  & 104  &   3   \\  
           & 4841      & 203   &  20  &   4   \\
           & 4850      & 262   &  30  &  15   \\   
           & 10450     & 106   &   4  &  15    \\ 
           & 10550     & 115   &   7  &  16, 18    \\  
%-----------------------------------------------------------------------------
           &           &       &      &       \\ 
J0200+4049 & 26.3      & 52780 &6000  &   9   \\ 
           & 74        & 5519  & 828  &    1  \\ 
           & 151       & 12080 & 846  &   10  \\ 
           & 178       & 2997  & 450  &   12  \\
           & 239       & 9129  &1369  &    P  \\
           & 325       & 2980  &  80  &   15  \\   
           & 327       & 4360  & 184  &   17  \\   
           & 333       & 10000 &1500  &    P  \\
           & 408       & 4390  & 219  &    3   \\ 
           & 605       & 5014  & 501  &    P  \\
           & 1400      & 292   &  29  &    2  \\     
           & 1400      & 220   &  10  &   17  \\   
           & 1400      & 1261  & 126  &    6  \\
           & 1460      & 1359  & 136  &    3  \\   
           & 4850      & 180   &  27  &    7  \\
           & 4850      & 292   &  35  &   15  \\  
           & 4850      & 418   &  10  &   17  \\  
           & 10450     &  95   &  11  &   15  \\   
           & 10550     &  95   &  2   &   17, 18  \\   
%-----------------------------------------------------------------------------
           &           &       &      &       \\ 
J0807+7400 & 38        &  3300 & 900 &    13  \\
           & 151       &  1220 &  90 &    14  \\
           & 239       &  929  & 139 &     P  \\
           & 334       &  937  & 141 &     P  \\
           & 605       &  433  &  43 &     P  \\ 
           & 1289      &  169  &  17 &     P  \\
           & 1400      &  155  &  16 &     2  \\ 
           & 1465      &  152  &  15 &     5  \\
           & 4850      &   46  &   7 &     8  \\     
           & 4850      &   38  &   6 &     7  \\     
%---------------------------------------------------------------------------------------------------------------------
\hline
\end{tabular}

P: Present paper
1: VLA Low-frequency Sky Survey (VLSS; {\tt http://lwa.nrl.navy.mil/VLSS}; 
2: NRAO VLA Sky Survey (NVSS; Condon et al. 1998);  
3: Vigotti et al. 1989; 
4: Konar et al. 2004; 
5: Lara et al. 2001;
6: White \& Becker 1992; 
7: Becker, White \& Edwards 1991; 
8: Gregory \& Condon 1991;       
9: Viner \& Erickson 1975;
10: Hales, Baldwin \& Warner 1993;
11: Pilkington \& Scott 1965;
12: Gower, Scott \& Wills 1967;
13: Hales et al. 1995;
14: Hales et al. 1991;
15. Schoenmakers et al. 2000;
16. Mack et al. 1994;
17. Vigotti et al. 1999;
18. Gregorini et al. 1998.  \\
\end{table}
%%%%%%%%%%%%%%%%%%%%%%%%%%%%%%%%%%%%%%%%%%%%%%%%%%%%%%%%%%%%%%%%%%%%%%%%%%%%%%%%%%%%%%%%%%%%%%%%%%%%%%%%%%%%%

\section{Discussion and results}
The integrated flux densities of the sources from our measurements as well as those
from the literature are listed in Table 3, which is self-explanatory, while 
some of the physical properties of the sources are listed in Table 4 which is
arranged as follows.  Column 1: source name;  Column 2:
redshift; columns 3 and 4: the largest angular size in arcsec and 
the corresponding projected linear size in kpc; 
column 5: the luminosity at 1.4 GHz in logarithmic units of W Hz$^{-1}$;
columns 6 and 10: lobe designation where W and E denote the western and eastern lobes respectively; 
columns 7 and 11: break frequency, $\nu_B$, with errors in units of GHz;
columns 8 and 12: magnetic field, B$_{eq}$, in units of nT; columns 9 and 13: spectral ages of the lobes 
with errors in units of Myr. 

For all the lobes listed in Table 2 we have fitted the observed spectra 
for the Jaffe \& Perola (1973, JP) model using the {\tt SYNAGE} package (Murgia 1996; Murgia et al. 1999) to
estimate the break frequency listed in columns 7 and 11 of Table 4.
Since the integrated spectra at low frequencies are usually determined
from a larger number of measurements than those of the individual lobes of emission, we have
fixed the value of $\alpha_{\rm inj}$ from a fit to the integrated spectrum
using the {\tt SYNAGE} package. However, because of the very diffuse extended
emission in these sources, several measurements even at low frequencies appear 
to have missed a significant amount of the total flux density. 
The highly discrepant measurements have not been used in
the final fits for estimating $\alpha_{\rm inj}$, where the typical error in 
$\alpha_{\rm inj}$ is within about $\pm$0.1.  For a sample of
GRSs studied earlier (Konar et al. 2008; Jamrozy et al. 2008), the values of $\alpha_{\rm inj}$
estimated for the lobes when possible are similar, within the uncertainties, to 
those estimated from the integrated spectra. Therefore this procedure has been adopted for
all the three sources considered here. We also examined the spectral index images of the
sources, but due to the large diffuse nature of the lobes and different uv coverages at 
different frequencies, age estimates using the integrated spectra of the lobes were found
to be more reliable. While estimating $\alpha_{\rm inj}$,
the core flux density has been subtracted from the total flux density.   

The  minimum energy magnetic field has been estimated by  
integrating the spectrum from a frequency corresponding to a
minimum Lorentz factor, $\gamma_{\rm min}$$\sim$10 for the relativistic electrons to an upper
limit of 100 GHz, which corresponds to a Lorentz factor ranging from a few times 10$^4$ to 10$^5$ 
depending on the estimated magnetic field strength 
(see Hardcastle et al. 2004; Croston et al. 2005; Konar et al. 2008).
Then, under assumptions that (i) the magnetic field strength in a given lobe is
constant throughout the energy-loss process, (ii) the particles injected into the
lobe have a constant power-law energy spectrum, and
(iii) the time-scale of isotropization of the pitch angles of the particles is short
compared with their radiative lifetime, the spectral age, $\tau_{\rm spec}$, has been 
estimated from

\begin{equation}
\tau_{\rm spec}=50.3\frac{B^{1/2}}{B^{2}+B^{2}_{\rm iC}}\left\{\nu_{\rm br}(1+z)\right\}
^{-1/2} [{\rm Myr}],
\end{equation}

\noindent
where $B_{\rm iC}$=0.318(1+$z$)$^{2}$ is the magnetic field strength equivalent to
the cosmic microwave background radiation. Here $B$, the magnetic field strength of the
lobes, and $B_{\rm iC}$ are expressed
in units of nT, $\nu_{\rm br}$ is the spectral break frequency in GHz above which the
radio spectrum steepens from the initial power-law spectral index, $\alpha_{\rm inj}$, 
given by S$\propto\nu^{-\alpha_{\rm inj}}$. 

%%%%%%%%%%%%%%%%%%%%%%%%%%%%%%%%%%%%%%%%%%%%%%%%%%%%%%%%%%%%%%%%%%%%%%%%%
\begin{figure*}
\hbox{
  \psfig{file=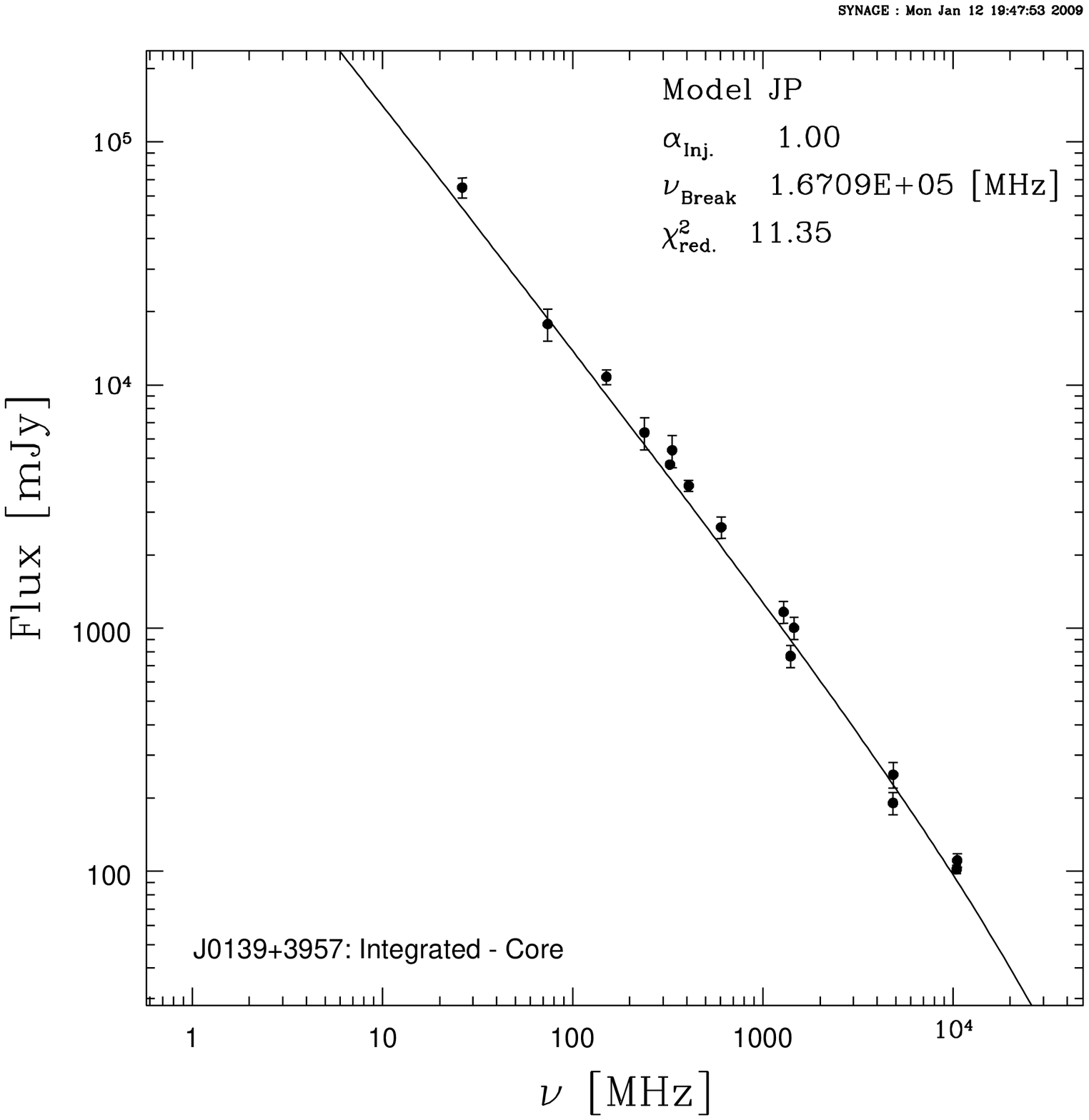,width=2.3in,angle=0}
  \psfig{file=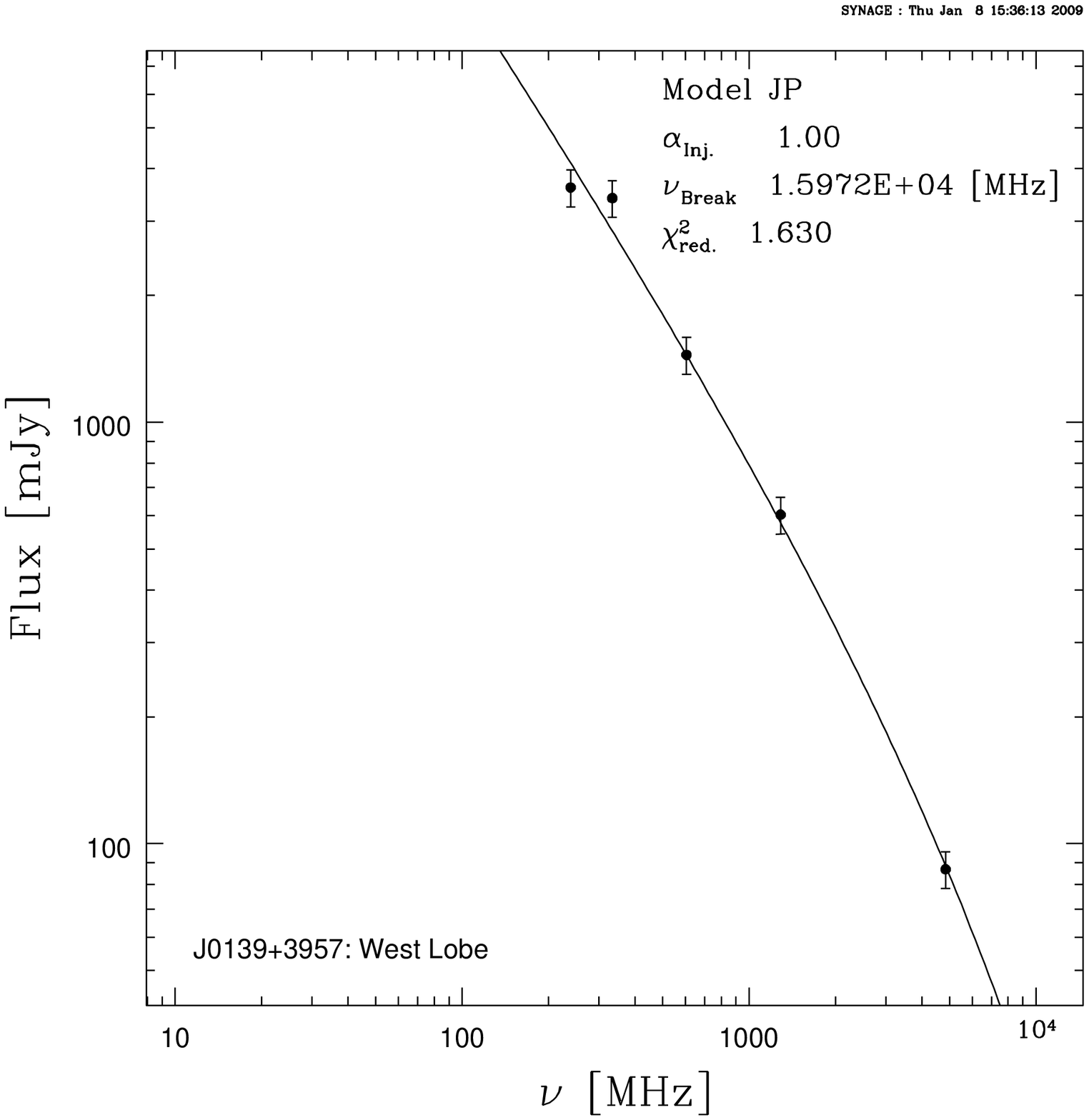,width=2.3in,angle=0}
  \psfig{file=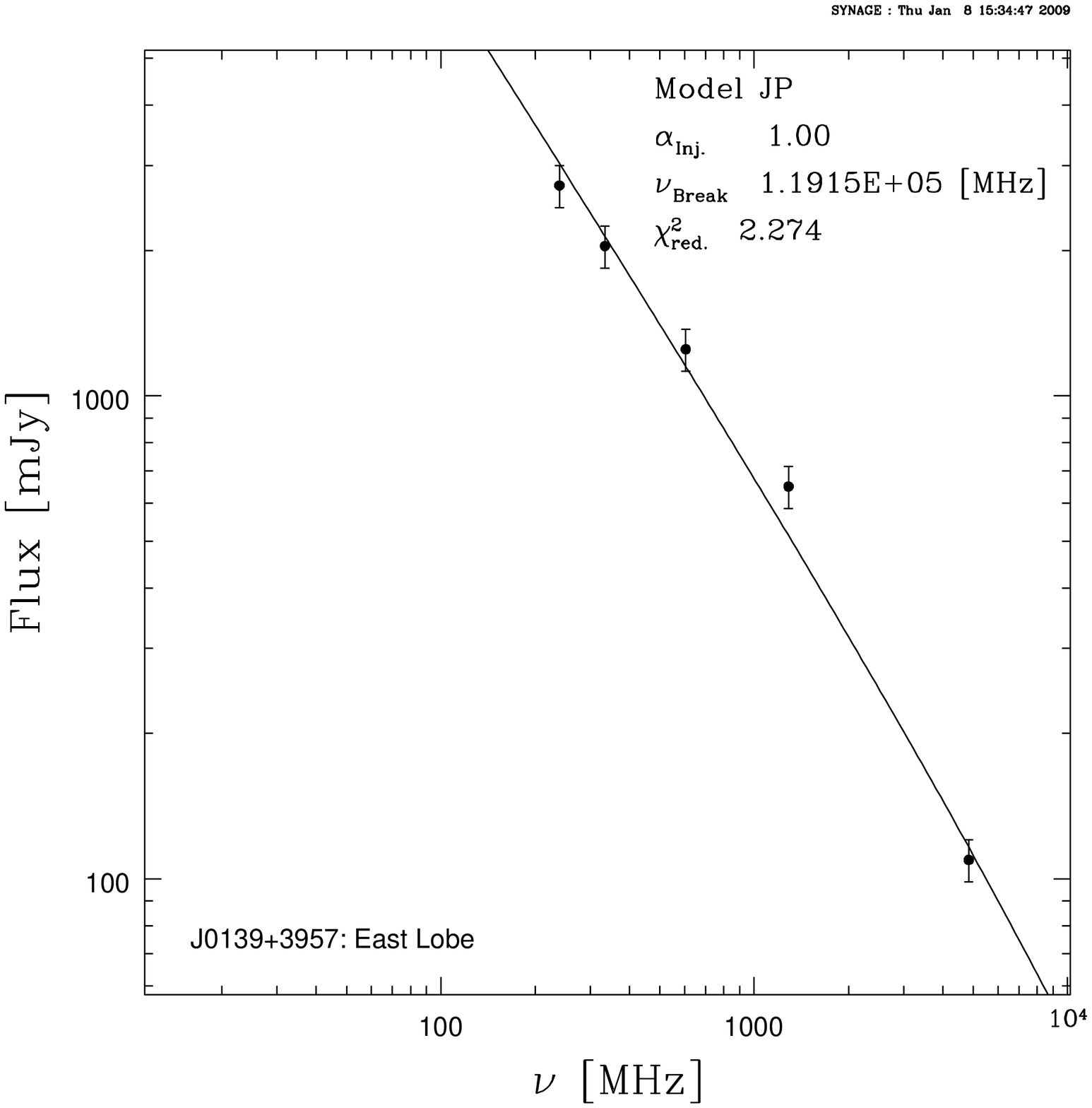,width=2.3in,angle=0}
}
\caption[]{The fits to the spectra of the entire source after subtracting the
contribution of the core (left panel), western (middle panel) and eastern (right panel)
lobes of J0139+3957 using the {\tt SYNAGE} package. 
           }
\end{figure*}
%%%%%%%%%%%%%%%%%%%%%%%%%%%%%%%%%%%%%%%%%%%%%%%%%%%%%%%%%%%%%%%%%%%%%%%%%

\subsection{J0139+3957}
The large-scale structure showing the
relaxed lobes (Fig. 1) has been reported earlier
by a number of authors (e.g. Hine 1979; Vigotti et al. 1989). 
Klein et al. (1995) presented Effelsberg observations of the source
and suggested it to be one with possible relic lobes of emission.
Konar et al. (2004)
reported GMRT and VLA observations at 1287 at 4841 MHz respectively and found 
that the spectral index of the entire source between
1.3 and 4.8 GHz is 1.34 while that of the core is 0.8. In Fig. 4, we show
the fits to the integrated spectra of the entire source as well as the western and 
eastern lobes using the injection spectral index, $\alpha_{\rm inj}$=1.00, 
determined from the fit to the entire source after subtracting the core flux
density. The spectral ages for the western and eastern lobes using our measurements 
have been found to be 12$^{+10}_{-1}$ and 
5.3$^{+29}_{-2.2}$  Myr respectively (Table 4). 

The existence of a possible steep-spectrum core 
was noted earlier (e.g. Hine 1979; Klein et al. 1995).
Similar-resolution observations of about 5 arcsec between 1.4 and 5 GHz
(Fomalont \& Bridle 1978; Gregorini et al. 1988; Bondi et al. 1993;
Klein et al. 1995) also yield a core spectral index of $\sim$0.8.  This
is consistent with the 10.6 GHz value of
4$\pm$1 mJy (Mack et al.  1994). Saripalli et al. (1997) find the core
to be a compact source with a flux density of 20$\pm$5 mJy from VLBI
observations at 1.67 GHz with a resolution of 25 mas. The core flux
densities from our low-frequency measurements, by putting a lower uv limit
to minimise any contamination from extended emission, along with values
from the literature are listed in 
Table 5. The spectrum (Fig. 5) shows clearly that 
while the high frequency spectrum is steep, it flattens below 
about a GHz. 

%%%%%%%%%%%%%%%%%%%%%%%%%%%%%%%%%%%%%%%%%%%%%%%%%%%%%%%%%%%%%%%%%%%%%%%%%
\begin{figure}
\hbox{
  \psfig{file=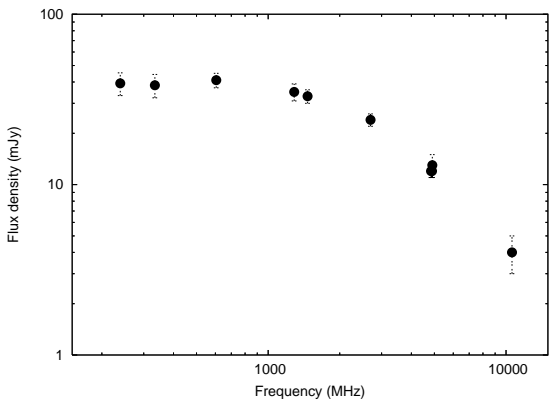,width=3.0in,angle=-90}
}
\caption[]{The spectrum of the radio core of J0139+3957.}
\end{figure}
%%%%%%%%%%%%%%%%%%%%%%%%%%%%%%%%%%%%%%%%%%%%%%%%%%%%%%%%%%%%%%%%%%%%%%%%%

%%%%%%%%%%%%%%%%%%%%%%%%%%%%%%%%%%%%%%%%%%%%%%%%%%%%%%%%%%%%%%%%%%%%%%%%%
\begin{figure*}
\hbox{
  \psfig{file=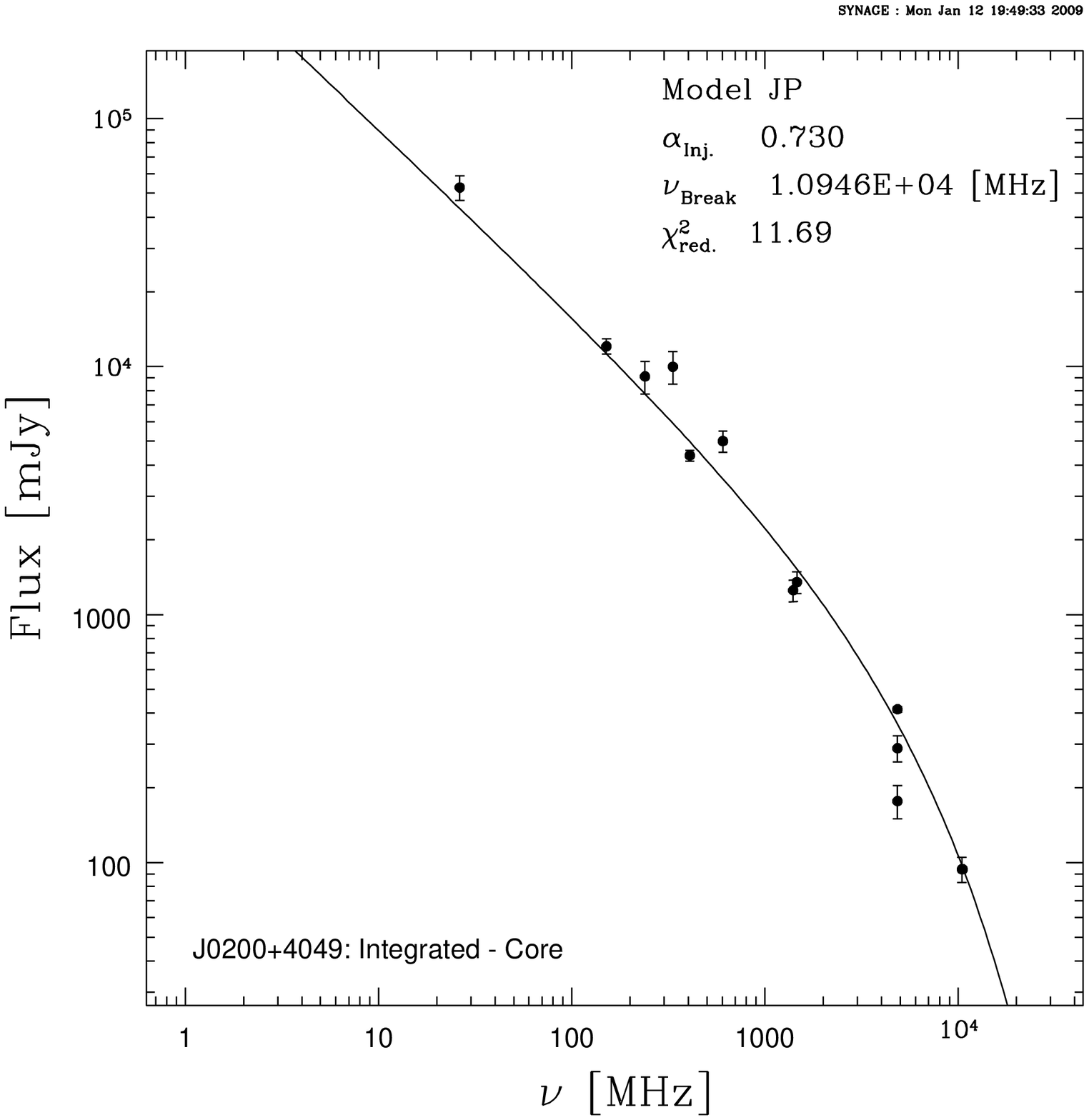,width=2.3in,angle=0}
  \psfig{file=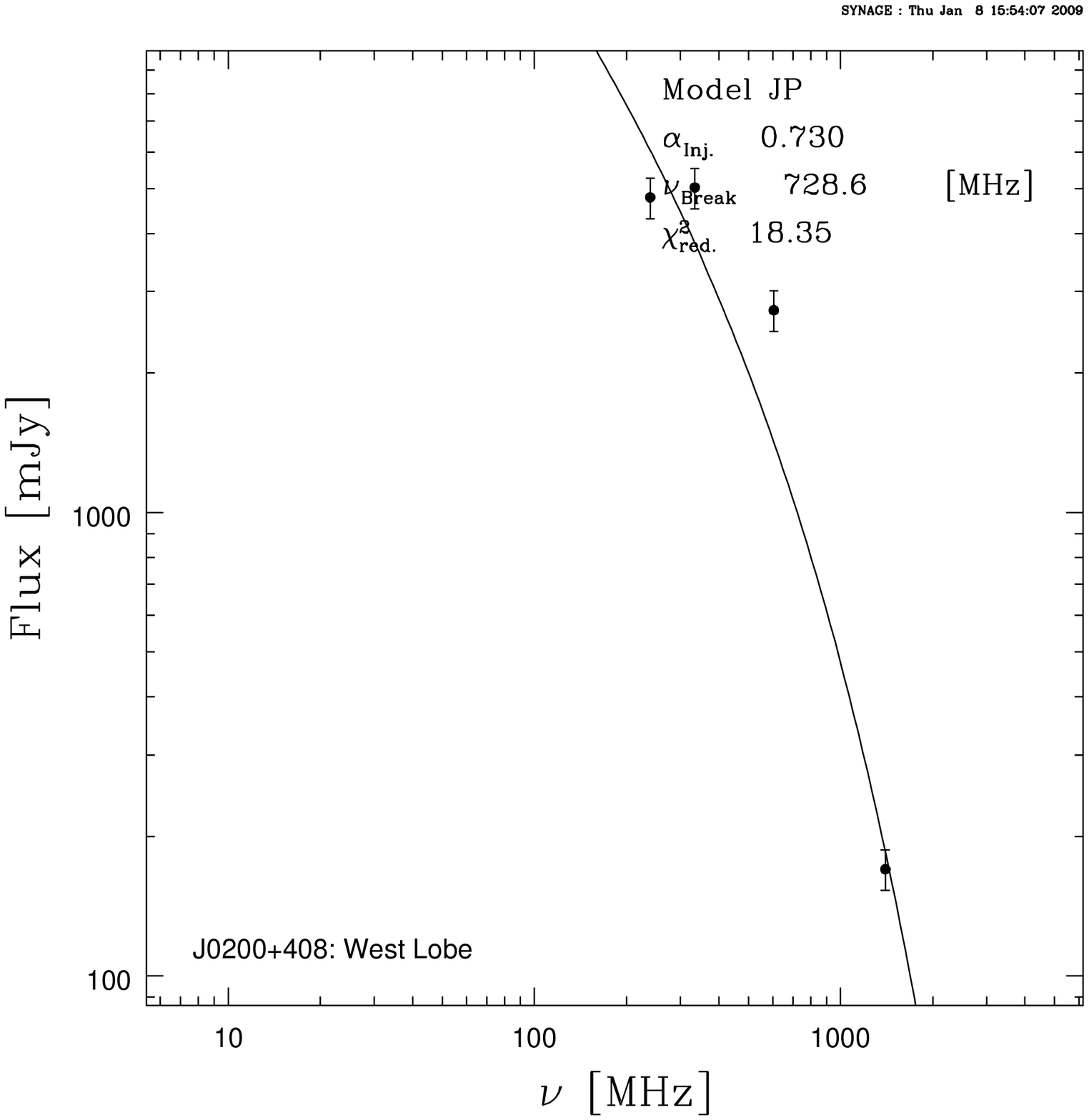,width=2.3in,angle=0}
  \psfig{file=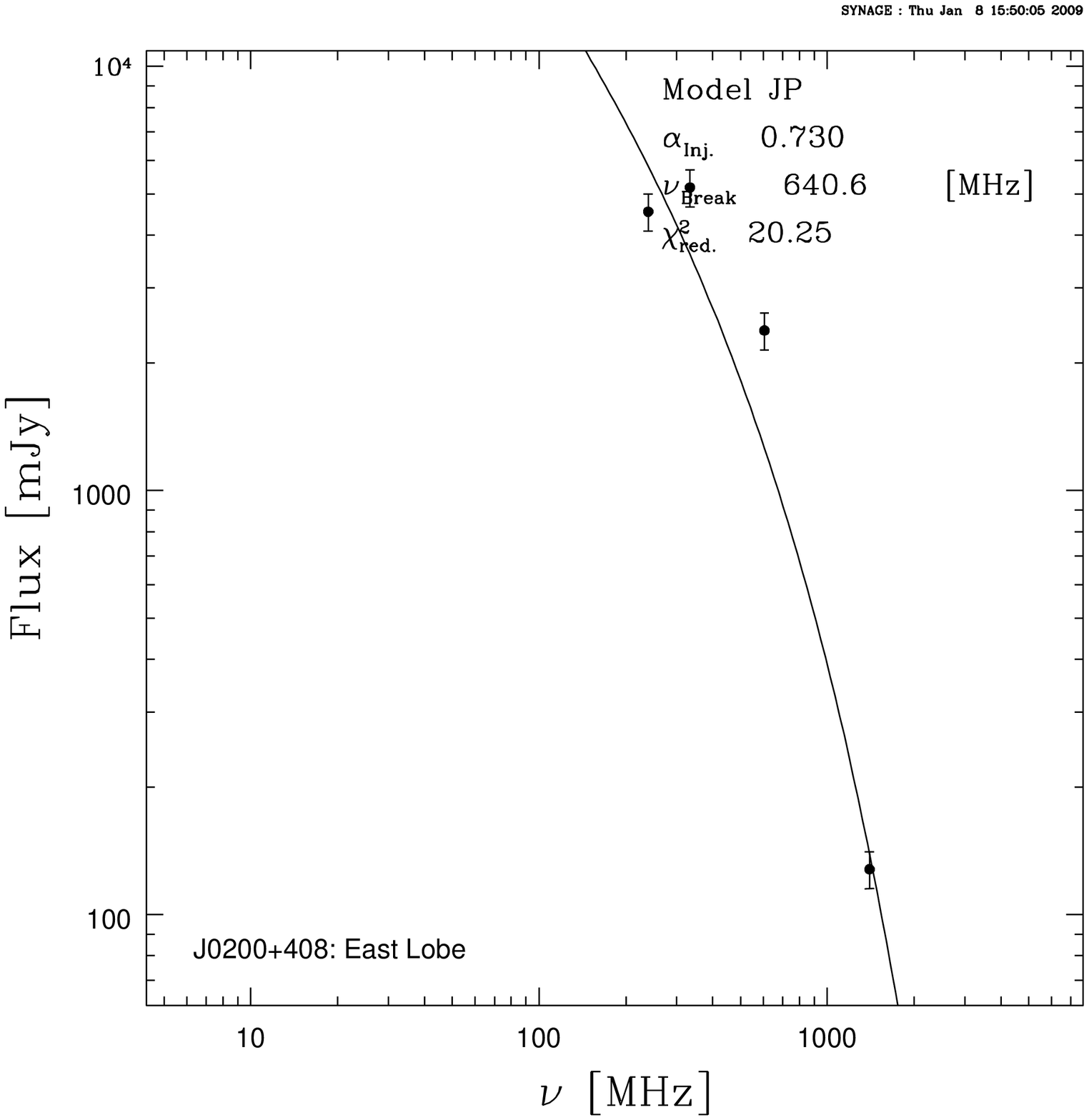,width=2.3in,angle=0}
}
\caption[]{The fits to the spectra of the entire source after subtracting the
contribution of the core (left panel), as well as the
western (middle panel) and eastern (right panel) lobes of J0200+4049 using the {\tt SYNAGE} package.
           }
\end{figure*}
%%%%%%%%%%%%%%%%%%%%%%%%%%%%%%%%%%%%%%%%%%%%%%%%%%%%%%%%%%%%%%%%%%%%%%%%%

%%%%%%%%%%%%%%%%%%%%%%%%%%%%%%%%%%%%%%%%%%%%%%%%%%%%%%%%%%%%%%%%%%%%%%%%%%%%%%%%%%%%%%%%%
\begin{table*}
\caption {Physical properties of the sources}

\begin{tabular}{l c l c r c r c r r c r r}
\hline
Source   & Redshift&LAS&{\it l}&P$_{1.4}$        & Cmp.&$\nu_B$                & B$_{eq}$&Spec.            & Cmp.& $\nu_B$                 & B$_{eq}$&Spec.            \\
         &         &      &       &              &     &                       &         &age              &     &                         &         &age              \\
         &  &$^{\prime\prime}$&kpc&W Hz$^{-1}$&        & GHz                   &   nT    & Myr             &     & GHz                     &   nT    & Myr             \\
   (1)     & (2)   & (3)  & (4)  & (5)  & (6)  & (7)  &(8)& (9)  & (10) & (11)  &   (12)  &(13)   \\
\hline                                                                                                                                         
J0139+3957&  0.2107  &  370 & 1259  & 26.16 &  W  &16$^{+4.3}_{-11.3}$    &  0.81   &12$^{+10}_{-1}$   & E   &119$^{+220}_{-116}$      &  0.66   & 5.3$^{+29}_{-2.2}$   \\
J0200+4049&  0.0827  &  920 & 1414  & 25.34 &  W  &0.73$^{+0.05}_{-0.20}$ &  0.20   &143$^{+25}_{-5}$ & E   &0.64$^{+0.04}_{-0.17}$   &  0.18   & 151$^{+25}_{-5}$   \\
J0807+7400&  0.1204  &  555 & 1190  & 24.76 &  W  &5.0$^{+27.7}_{-4.0}$   &  0.15   & 46$^{+57}_{-28}$& E   &2.8$^{+5.6}_{-1.3}$      &  0.20   & 64$^{+24}_{-27}$  \\
\hline

\end{tabular}
\end{table*}  
%%%%%%%%%%%%%%%%%%%%%%%%%%%%%%%%%%%%%%%%%%%%%%%%%%%%%%%%%%%%%%%%%%%%%%%%%

%%%%%%%%%%%%%%%%%%%%%%%%%%%%%%%%%%%%%%%%%%%%%%%%%%%%%%%%%%%%%%%%%%%%%%%%%
\begin{table*}
\caption{Core flux densities of J0139+3957}
\begin{tabular}{l r l r r r}
\hline
 Teles-    &  Resolution                               &     Date   & Freq.   &  S            & Ref.   \\
 cope      &  $^{\prime\prime}$                        &            & MHz     &  mJy          &        \\
  (1)      &     (2)                                   &   (3)      &  (4)    &  (5)          & (6)    \\
\hline
GMRT       &  $\sim$10                                 &2007 Jun 02 & 239     & 39.3$\pm$6 & P     \\
GMRT       &  $\sim$10                                 &2005 Jan 28 & 334     & 38.3$\pm$6 & P     \\
GMRT       &  $\sim$5                                  &2007 Jun 02 & 605     & 41.0$\pm$4 & P     \\
GMRT       &  $\sim$5                                  &2003 Aug 19 &1287     & 35.0$\pm$4 & 2     \\
VLA        &  $\sim$5                                  &1982 Sep    &1465     & 33.0$\pm$3 & 1     \\
Cambridge  &  $\sim$5                                  &1977 Oct    &2695     & 24.0$\pm$2 & 5     \\
VLA        &  $\sim$15                                 &2000 Jul 24 &4841     & 12.0$\pm$1 & 2     \\
VLA        &  4.9                                      &1989 Jul    &4885     & 12.0$\pm$1 & 4     \\
VLA        &  $\sim$2                                  &1977 Mar-May&4900     & 13.0$\pm$2 & 6     \\
Effelsberg &  69                                       &1990$-$1991 &10550    &  4.0$\pm$1 & 3     \\  
\hline
\end{tabular}

P: Present paper; 1: Gregorini et al. 1988; 2: Konar et al. 2004; 3: Mack et al. 1994;
4: Bondi et al. 1993; 5: Hine 1979; \\ 6: Fomalont \& Bridle 1978
\end{table*}
%%%%%%%%%%%%%%%%%%%%%%%%%%%%%%%%%%%%%%%%%%%%%%%%%%%%%%%%%%%%%%%%%%%%%%%%%

%%%%%%%%%%%%%%%%%%%%%%%%%%%%%%%%%%%%%%%%%%%%%%%%%%%
\begin{table*}
\caption{Core flux densities of J0200+4049}
\begin{tabular}{l r l r r r}
\hline
 Teles-    &  Resolution                               &     Date   & Freq.   &  S            & Ref.   \\
 cope      &  $^{\prime\prime}$                        &            & MHz     &  mJy          &        \\
  (1)      &     (2)                                   &   (3)      &  (4)    &  (5)          & (6)    \\
\hline

GMRT       &  $\sim$11                                 &2007 Dec 29 & 239     & 7.6 $\pm$1.1  & P     \\
GMRT       &  $\sim$7                                  &2004 Dec 25 & 333     & 9.5 $\pm$1.4  & P     \\
GMRT       &  $\sim$5                                  &2007 Dec 29 & 605     & 9.2 $\pm$0.9  & P     \\
GMRT       &  $\sim$3                                  &2004 Nov 25 &1289     & 7.9 $\pm$0.8  & P     \\
VLA        &  $\sim$12                                 &2000 Jul 24 &4841     & 2.6 $\pm$0.3  & 1     \\
\hline
\end{tabular}

P: Present paper; 1: Konar et al. 2004
\end{table*}
%%%%%%%%%%%%%%%%%%%%%%%%%%%%%%%%%%%%%%%%%%%%%%%%%%%

%%%%%%%%%%%%%%%%%%%%%%%
\subsection{J0200+4049}
The double lobed structure has been imaged earlier by a number of
authors (Vigotti et al. 1989; Gregorini et al. 1988;
Schoenmakers et al. 2000), and some of the diffuse emission is also
visible in the NVSS image reproduced by Konar et al. (2004). 
The latter also report the detection of a radio core at 4.8 GHz 
with a flux density of 2.6 mJy, and a compact component to the east which
is coincident with a faint galaxy seen in the Digital Sky
Survey (DSS) image. Our images (Fig. 2) show evidence of a depression
in the centre of the western lobe, which is clearly seen in the 333-MHz
image. It is also seen at 239 MHz where the data is of poorer quality,
and at 605 MHz which require more short-spacing data to produce a better
image. The GMRT image at 1289 MHz is not shown here since only the core
and the unrelated compact source to the east (see Konar et al. 2004) were  
visible.

The flux density of the source appears to have been under-estimated in
the 74-MHz VLSS image and at 178 MHz in the 4C survey, possibly due to
the large extent of the diffuse low-brightness lobes of emission. The
values at these frequencies lie significantly
below the fit to the integrated spectrum (Fig. 6). For example the flux density at
178 MHz is only $\sim$3 Jy while it has been estimated to be
approximately 12 Jy at 151 MHz. The flux density values at 74 and 178 MHz 
have not been considered in the fit to the integrated spectrum of the source.  
The injection spectral index has been estimated to be 0.73. 
The spectral ages estimated for the western and eastern lobes using
our measurements are 
143$^{+25}_{-5}$ and 151$^{+25}_{-5}$  Myr respectively (Table 4; Fig. 6), 
which are the largest amongst the three sources discussed here. More accurate
measurements of the flux density at both low and high frequencies would help
determine the break frequency more reliably. The lobes are 
relaxed with a depression in surface brightness in the western lobe. This
clearly suggests that the lobes are no longer being fed with energy from 
the nucleus. 

The core flux densities estimated by imaging with lower uv limits to
minimise contamination by extended emission are listed in Table 6.
Although the high-frequency spectrum appears to be fairly steep, with a 
spectral index of $\sim$0.8 between 1289 and 4841 MHz, the spectrum appears
to turn over at low frequencies (Fig. 7).

%%%%%%%%%%%%%%%%%%%%%%%%%%%%%%%%%%%%%%%%%%%%%%%%%%%%%%%%%%%%%%%%%%%%%%%%%%
\begin{figure}
\hbox{
  \psfig{file=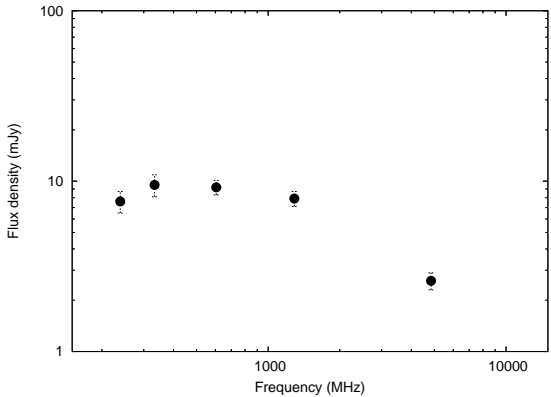,width=3.0in,angle=-90}
}
\caption[]{The spectrum of the radio core of J0200+4049.}
\end{figure}
%%%%%%%%%%%%%%%%%%%%%%%%%%%%%%%%%%%%%%%%%%%%%%%%%%%%%%%%%%%%%%%%%%%%%%%%%
%
%%%%%%%%%%%%%%%%%%%%%%%%%%%%%%%%%%%%%%%%%%%%%%%%%%%%%%%%%%%%%%%%%%%%%%%%%%
\begin{figure*}
\hbox{
  \psfig{file=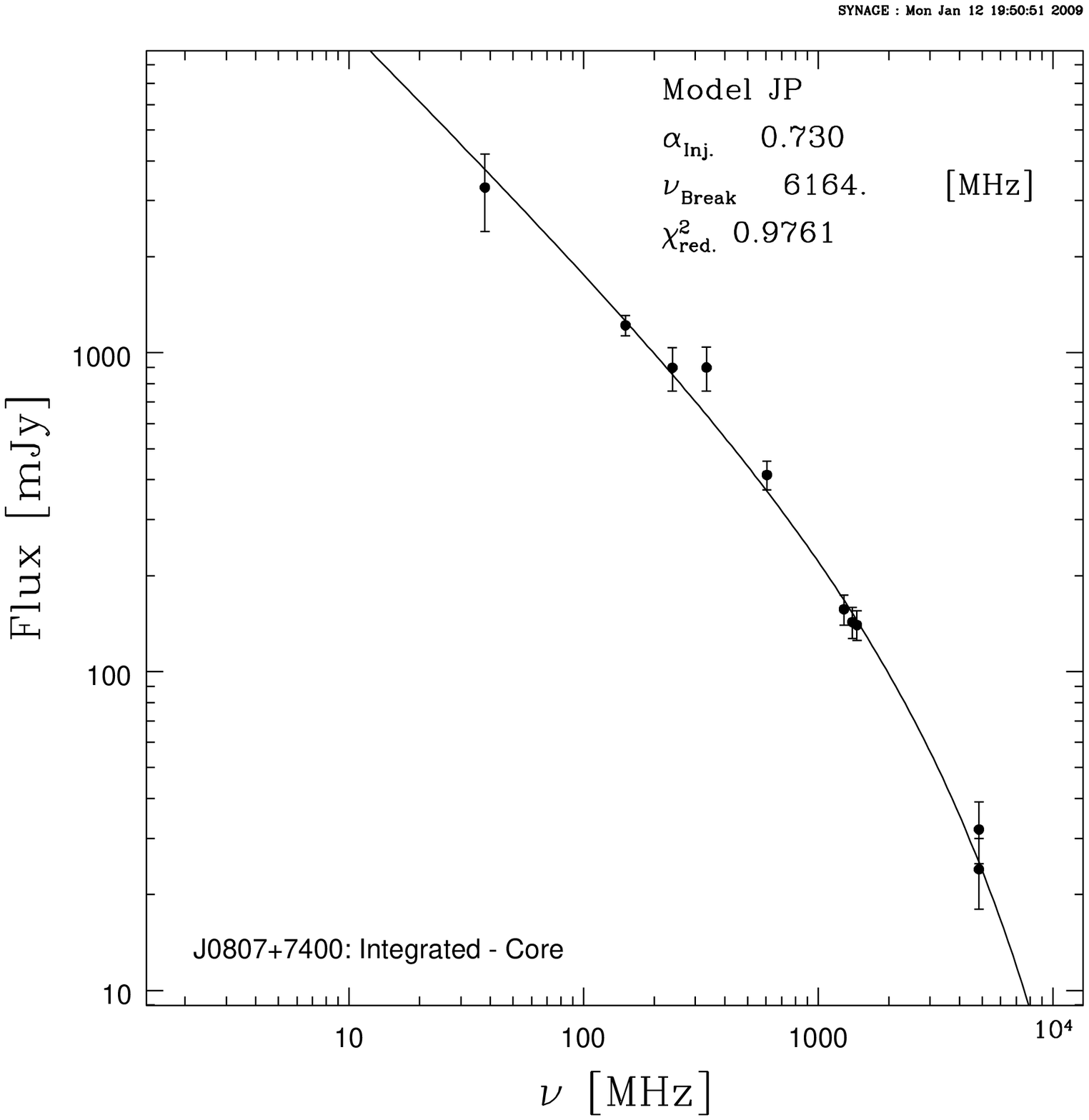,width=2.3in,angle=0}
  \psfig{file=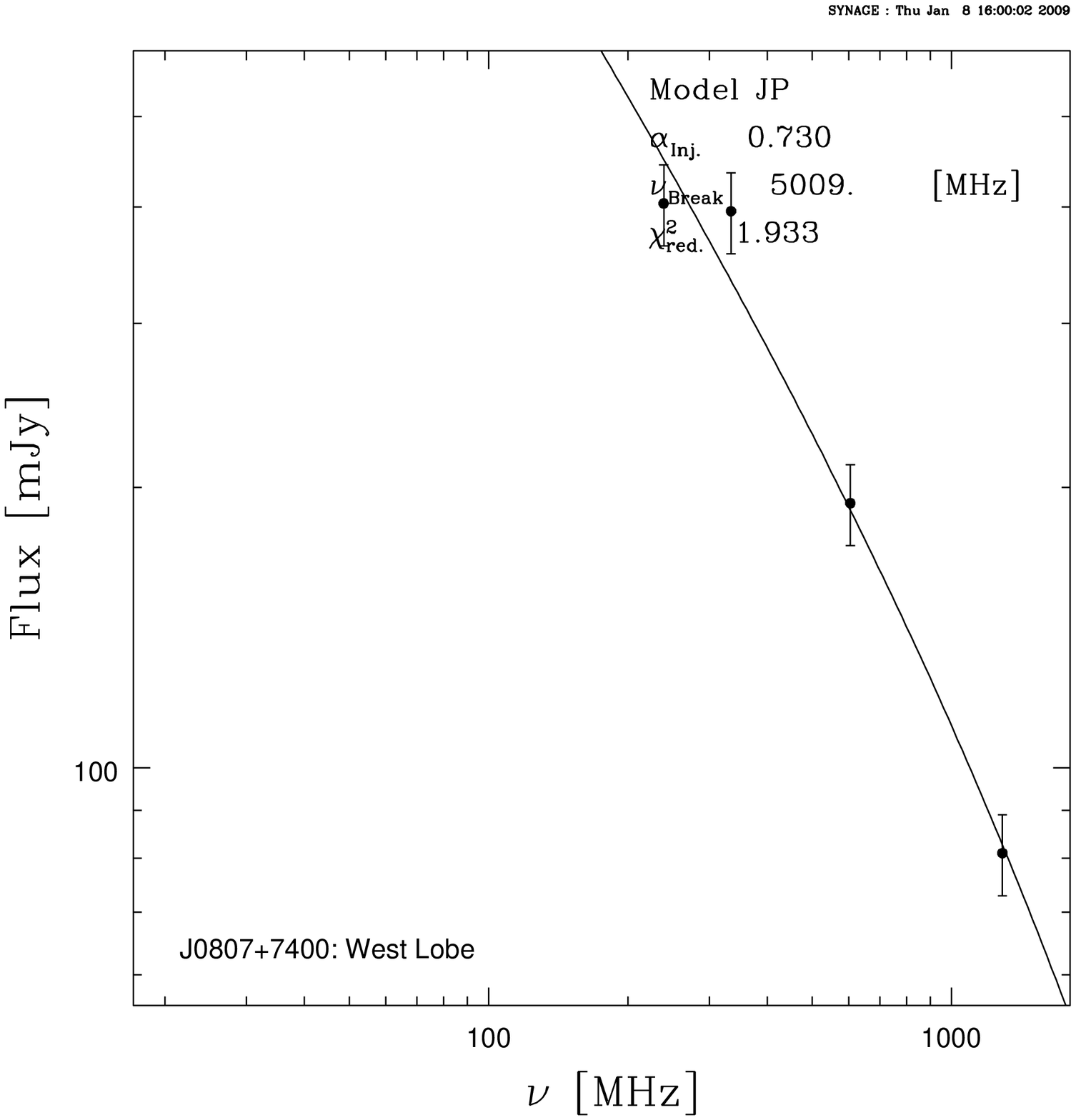,width=2.3in,angle=0}
  \psfig{file=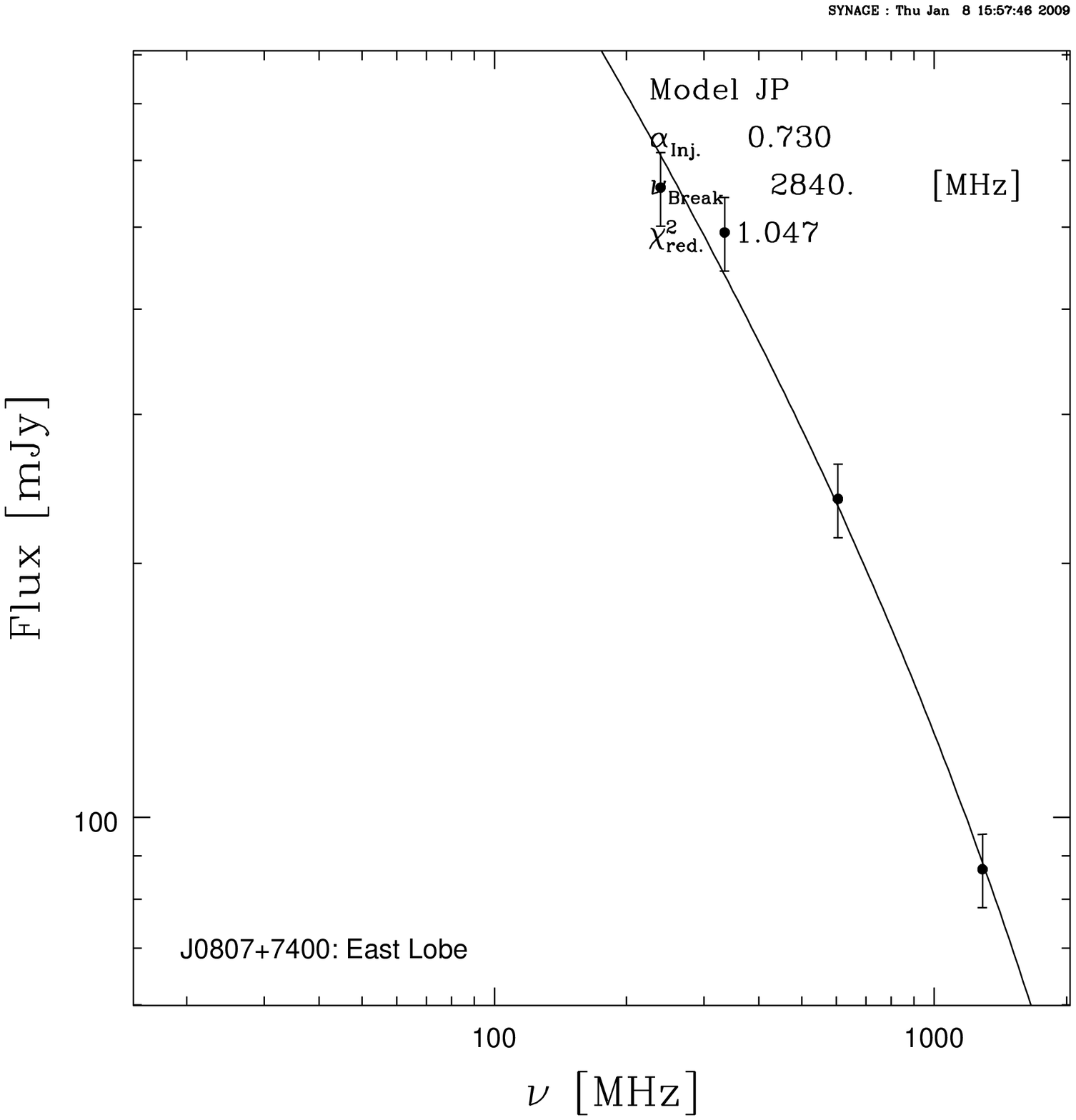,width=2.3in,angle=0}
}
\caption[]{The fits to the spectra of the entire source after subtracting the
contribution of the core (left panel) as well as the 
western (middle panel) and eastern (right panel) lobes of J0807+7400 using the {\tt SYNAGE} package.
           }
\end{figure*}
%%%%%%%%%%%%%%%%%%%%%%%%%%%%%%%%%%%%%%%%%%%%%%%%%%%%%%%%%%%%%%%%%%%%%%%%%

%%%%%%%%%%%%%%%%%%%%%%%%
\subsection{J0807+7400}
This source is a giant low-power radio galaxy, the weakest of our three
GRSs. Observations at 1.4 GHz by Lara et al. (2001)  show a compact
core component and a weak and extended halo-like emission elongated in the
east-west direction with no evidence of either jets or hotspots. At 4.9 GHz,
only the core component was visible. Lara et al. (2001) suggested that this object
could be a relic FR II radio galaxy where hotspot regions are no longer present.
Our GMRT images at low frequencies show the diffuse lobes and bridge of emission.
The spectral ages of the lobes using our measurements
are 46$^{+57}_{-28}$ and 64$^{+24}_{-27}$ Myr for the western and
eastern lobes respectively (Table 4; Fig. 8).

The core has a flat spectral index at high frequencies but appears to
have a steep radio spectrum with a spectral index of $\sim$0.6 at lower frequencies
(Fig. 9), possibly due to unresolved jet/lobe structure from more recent activity.

%%%%%%%%%%%%%%%%%%%%%%%%%%%%%%%%%%%%%%%%%%%%%%%%%%%%%%%%%%%%%%%%%%%%%%%%%%%%
\begin{table*}
\caption{Core flux densities of J0807+7400}
\begin{tabular}{l r l r r r}
\hline
 Teles-    &  Resolution                               &     Date   & Freq.   &  S            & Ref.   \\
 cope      &  $^{\prime\prime}$                        &            & MHz     &  mJy          &        \\
  (1)      &     (2)                                   &   (3)      &  (4)    &  (5)          & (6)    \\
\hline

GMRT       &  $\sim$13                                 &2005 Jan 07 & 239     & 32.0$\pm$5  & P     \\
GMRT       &  $\sim$9                                  &2004 Dec 07 & 334     & 38.4$\pm$6  & P     \\
GMRT       &  $\sim$5                                  &2005 Jan 07 & 605     & 19.1$\pm$2  & P     \\
GMRT       &  $\sim$3                                  &2005 Jan 17 &1289     & 12.5$\pm$1  & P     \\
VLA-B      &  $\sim$6                                  &1995 Nov 19 &1385     & 13.4$\pm$1  & P     \\
VLA-C      &  $\sim$14                                 &1996 Feb 19 &1685     & 10.9$\pm$1  & P     \\
VLA-D      &  $\sim$45                                 &1993 Nov 01 &1435     & 10.5$\pm$1  & P     \\
VLA-B+C    &  $\sim$12                                 &1995 Nov 19 &1465     & 14.0$\pm$1  & 1     \\
           &                                           &1996 Feb 19 &1465     &                       \\
VLA-C      &  $\sim$6                                  &1996 Feb 19 &4860     & 13.5$\pm$1  & P     \\
\hline
\end{tabular}

P: Present paper; 1: Lara et al. 2001
\end{table*}
%%%%%%%%%%%%%%%%%%%%%%%%%%%%%%%%%%%%%%%%%%%%%%%%%%%%%%%%%%%%%%%%%%%%%%%%%%%%
%

\section{Concluding remarks}
We have presented low-frequency GMRT observations of the lobes of emission 
of three giant radio sources, J0139+3957, J0200+4049 and J0807+7400, with diffuse 
lobes of emission, but no hotspots and no jets from the nuclear region. Klein et al.
(1995) had earlier suggested J0139+3957 to consist of relic lobes, while Lara et al.
(2001) suggested that J0807+7400 is a relic FRII radio source. We have tried to explore
this theme by examining the structure and spectra of all three sources 
over a large frequency range. Although these three sources do not have  
hotspots, their structures are not similar to the FRI sources which are 
characterised by jets that expand to form the diffuse lobes of emission. 
Also, the luminosities of two of the three sources are above the dividing 
line for these two classes of sources.

The spectral ages of the lobes estimated from {\tt SYNAGE} fits to the spectra of
the lobes, as well as their integrated spectra,
are in the range of $\sim$5$\times$10$^6$ to 1.5$\times$10$^8$ yr, the upper range being
close to time scales for which the lobes are likely to remain visible if not fed with
a fresh supply of energy from the parent galaxy. 

%%%%%%%%%%%%%%%%%%%%%%%%%%%%%%%%%%%%%%%%%%%%%%%%%%%%%%%%%%%%%%%%%%%%%%%%%
\begin{figure}
\hbox{
  \psfig{file=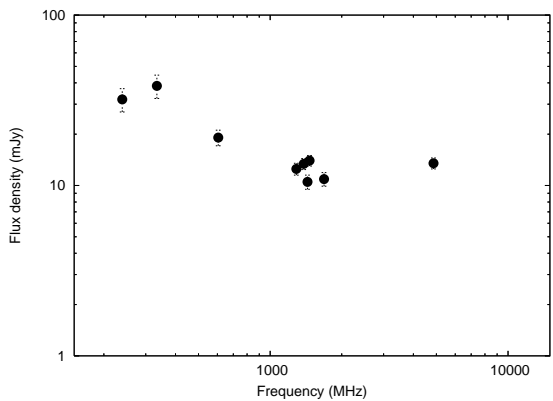,width=3.0in,angle=-90}
}
\caption[]{The spectrum of the radio core of J0807+7400.}
\end{figure}
%%%%%%%%%%%%%%%%%%%%%%%%%%%%%%%%%%%%%%%%%%%%%%%%%%%%%%%%%%%%%%%%%%%%%%%%%

The structure and spectra suggest that these lobes are possibly no longer being fed,
with one of the lobes in J0200+4049 exhibiting a depression in surface brightness
towards the centre of the lobe. 
However, all three have detected radio cores, two of which, J0139+3957 and J0200+4049,  
have spectra which flatten at lower frequencies, which could be indicative 
of an active core, while the core spectrum of J0807+7400 appears steep
at low frequencies suggesting unresolved extended emission.  The detection of cores
suggests that their nuclei are currently active. Although these sources do not belong
to the classic double-double radio galaxies, about a dozen or so of which are presently
known (e.g. Saikia, Konar \& Kulkarni 2006), these might also represent sources with
evidence of episodic nuclear activity.  It would be interesting to determine the 
structures of the cores from high-resolution radio observations.  The relationship
between the turnover frequency and source size for GPS objects (O'Dea 1998) shows that
the source sizes are $\sim$0.5 to 2 kpc for turnover frequencies of
$\sim$1 and 0.4 GHz respectively (see Figs. 5 and 7).  Interpreting the 
cores as more recent activity, the time scales of episodic activity would range from
$\sim$10$^7$ to 10$^8$ yr.  
  
%%%%%%%%%%%%%%%%%%%%%%%%%%%%%%%%%%%%%%%%%%%%%%%%%%%%%%%%%%%%%%%%%%%%%%%%%
\section*{Acknowledgments}
We thank an anonymous referee for helpful comments.
The Giant Metrewave Radio Telescope is a national facility operated by the 
National Centre 
for Radio Astrophysics of the Tata Institute of Fundamental Research. We thank the staff
for help with the observations.  The National Radio Astronomy Observatory  is a
facility of the National Science Foundation operated under co-operative
agreement by Associated Universities Inc. We thank the VLA staff for easy access
to the archival data base.  This research has made use of the NASA/IPAC extragalactic database (NED)
which is operated by the Jet Propulsion Laboratory, Caltech, under contract
with the National Aeronautics and Space Administration.  We thank numerous contributors
to the GNU/Linux group.

{}

\end{document}